\documentclass[aps,prb,reprint,showpacs,superscriptaddress]{revtex4-2}
\usepackage{graphicx}
\usepackage[hidelinks]{hyperref}
\usepackage{amsmath,bm,amssymb}
\usepackage{float}
\usepackage{subfigure}
\usepackage[english]{babel}
\usepackage{comment}

\newcommand{\li}{\begin{aligned}}
\newcommand{\eli}{\end{aligned}}

\newcommand{\be}{\begin{equation}}
\newcommand{\ee}{\end{equation}}

\begin{document}

\title{Generalized continuum theory of phonon angular momentum in crystals}

\author{Mamoru Matsuo}
\email{mamoru@ucas.ac.cn}
\affiliation{Kavli Institute for Theoretical Sciences, University of Chinese Academy of Sciences, Beijing, 100190, China}
\affiliation{CAS Center for Excellence in Topological Quantum Computation, University of Chinese Academy of Sciences, Beijing 100190, China}
\affiliation{Advanced Science Research Center, Japan Atomic Energy Agency, Tokai, 319-1195, Japan}
\affiliation{RIKEN Center for Emergent Matter Science (CEMS), Wako, Saitama 351-0198, Japan}

\author{Naoki Nishimura}
\affiliation{Institute for Solid State Physics, University of Tokyo, Kashiwa, 277-8581, Japan}

\author{Ai Yamakage}
\affiliation{Department of Physics, Nagoya University, Nagoya 464-8602, Japan}

\author{Takeo Kato}
\affiliation{Institute for Solid State Physics, University of Tokyo, Kashiwa, 277-8581, Japan}

\date{\today}

\begin{abstract}
We formulate a generalized continuum theory of phonon angular momentum in crystals by introducing a local $\mathrm{SO}(3)$ material frame in addition to the macroscopic displacement field.
The local frame represents rotational optical degrees of freedom of the unit cell and brings acoustic displacement modes and optical rotational modes into a common long-wavelength continuum description.
In the linearized limit, the co-rotated deformation gradient and the rotational gradient associated with the local material frame recover the Eringen microdeformation and wryness tensors; isotropic micropolar elasticity then appears as a special case.
Rotational symmetry and Noether's theorem determine the continuum phonon angular-momentum density, including both the displacement-polarization contribution and the intrinsic microrotation contribution.
The theory further identifies the locking limit in which microrotation reduces to lattice vorticity and the improper-symmetry-breaking terms responsible for chiral phonon splitting.
\end{abstract}

\maketitle

\section{Introduction}

The interplay between angular momentum and lattice dynamics in solids has attracted increasing attention over the past decade, following the introduction of phonon angular momentum~\cite{ZhangNiu2014}. Phonon angular momentum is conventionally defined as ${\bm S}_{\rm ph} = \sum_j M_j\, {\bm u}_j \times \dot{\bm u}_j$~\cite{Vonsovskii1962,McLellan1988,Garanin2015,Nakane2018,Ruckriegel2020}, where $M_j$ and ${\bm u}_j$ denote the mass and displacement of the $j$th atom. It is then expected that phonon angular momentum can be converted into other forms of angular momentum, such as rigid-body rotation or electron spin. Several theoretical studies have employed it as an indicator of the ability of lattice motion to generate rotation or spin polarization~\cite{Hamada2018,Park2020,Hamada2020,Ohe2024}.

To use angular momentum as a well-defined physical quantity, however, its meaning must be specified beyond doubly degenerate transverse acoustic modes~\cite{Vonsovskii1962,McLellan1988}. Its definition for optical phonons or short-wavelength phonons remains unsettled, and the relation between structural chirality and phonon angular momentum is likewise not fully understood. Phonons in chiral crystals lacking improper rotational symmetries are termed \textit{chiral phonons}~\cite{Juraschek2025}. Although the two transverse acoustic modes of chiral phonons share the same sound velocity, their dispersions split at finite wavenumber according to chirality~\cite{Hansen1980,Komiyama2022,Tsunetsugu2023,Kato2023,Tsunetsugu2026}.
For such acoustic chiral phonons, the conventional definition of phonon angular momentum is no longer sufficient.

A continuum definition of phonon angular momentum is needed to analyze gyromagnetic effects~\cite{EdH1915,Barnett1915,Scott1962}, which have regained attention in spintronics~\cite{Matsuo2011,Matsuo2013,Wallis-APL-2006-09,Zolfagharkhani2008,ono2015Barnett,ogata2017Gyroscopic,hirohata2018magneto,imai2018observation,imai2019angular,chudo2014Observation,chudo2015Rotational,harii2015Line,chudo2021Barnett,chudo2021Observation,chudo2025mechanical,wood2017magnetic,wood2018Sci.Adv.,jin2024NatCommun,Kobayashi2017,kurimune2020Highly,kurimune2020Observation,tateno2020Electrical,tateno2021Einstein,Harii-2019-NatCommun,Mori2020,okano2019nonreciprocal,horaguchi2025nanometer,takahashi2016Spin,takahashi2020Giant,tabaeikazerooni2020Electron,tabaeikazerooni2021Electrical,tokoro2022spin} and in quark-gluon plasma physics~\cite{adamczyk2017global,adam2018global,adam2019polarization,acharya2020evidence,adam2021global}.
In current applications, the gyromagnetic effect has been extended to local lattice rotation generated by surface acoustic waves~\cite{Kobayashi2017,kurimune2020Highly,kurimune2020Observation,tateno2020Electrical,tateno2021Einstein}
and by phonons~\cite{Funato2024}. In these settings, electron spins are assumed to couple to the local orbital motion of the lattice, rigidly locked to the vorticity, namely, the curl of the lattice velocity. Although this vorticity-based picture accounts for many spin-mechanical phenomena, it is limited to acoustic degrees of freedom and does not determine how electron spins couple to local lattice rotation when optical phonons are present.

A key limitation of existing continuum approaches is the absence of a common language in which acoustic displacement modes and optical rotational modes can be assigned angular momentum within the same conservation law.
This paper formulates such a continuum theory by introducing a local $\mathrm{SO}(3)$ material frame attached to the rotating microstructure, rather than by taking standard micropolar elasticity (MPE) itself as the main result.
The construction is compatible with the broader hierarchy of Eringen microcontinua~\cite{Eringen1999}, while the rotational sector considered here reduces to micropolar elasticity after linearization through the microdeformation and wryness tensors.
The central outcome is that rotational symmetry and Noether's theorem determine a definite continuum angular-momentum density, consisting of the displacement-polarization contribution and the intrinsic contribution from the local material rotation.
The same construction shows how the Cauchy limit emerges when microrotation locks to macroscopic vorticity and how improper-symmetry-breaking terms split circularly polarized phonons in chiral media.
The local material frame identifies the continuum variables needed for spin-rotation coupling in the presence of optical phonons, a coupling to be treated separately.
Refs.~\cite{Pouget1986a,Pouget1986b,Kishine2020} already pointed to several aspects of this physics; here we formulate a continuum definition of phonon angular momentum that keeps its relation to vorticity locking and chiral splitting explicit.

\section{Outline}

\subsection{MPE theory}
\label{MPEtheory}

We first review MPE, following Ref.~\cite{Kishine2020}.
Consider a continuum medium with displacement field ${\bm u}=(u_1,u_2,u_3)$ defined at position ${\bm x}=(x_1,x_2,x_3)$.
Throughout this paper,  Latin indices ($i, j, k, \dots = 1, 2, 3$) denote spatial components in  three-dimensional Euclidean space. The Einstein summation convention is implied for repeated indices, where summation is taken using the Euclidean metric $\delta_{ij}$ (e.g., $u_i u_i \equiv \sum_{i=1}^3 u_i u_i$). The Levi-Civita symbol is denoted by $\varepsilon_{ijk}$, and the spatial derivatives are abbreviated as $\partial_j \equiv \partial/\partial x_j$.

In conventional elasticity theory~\cite{LandauLifshitz1986}, the microdeformation is described by Cauchy's strain tensor $\epsilon_{ij}^{\rm C} = (\partial_i u_j + \partial_j u_i)/2$, which is a symmetric second-rank tensor.
In addition, the local rotation of the lattice is described by the axial rotation vector $\omega_i = \varepsilon_{ijk} \partial_j u_k/2$ (${\bm \omega} = {\rm rot} \, {\bm u}/2$).
Using the identity $\partial_i u_j = \epsilon_{ij}^{\rm C} + \varepsilon_{ijk} \omega_k$, one obtains
\begin{align}
    \epsilon_{ij}^{\rm C} = \partial_i u_j - \varepsilon_{ijk} \omega_k.
    \label{epsilonijC}
\end{align}
Conventional elasticity thus keeps local translations of material points and the associated force stress (force per unit area), while independent rotations and the corresponding couple stress are absent.

\begin{figure}
    \centering    \includegraphics[width=0.8\columnwidth]{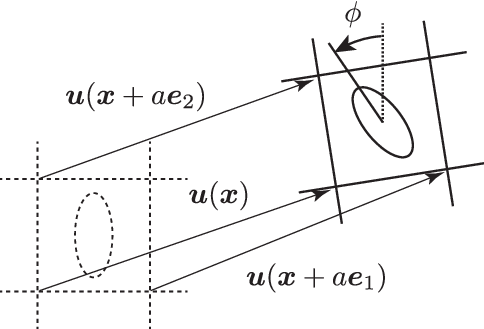}
    \caption{Two-dimensional schematic of the microrotation field ${\bm \phi}$ and the displacement field ${\bm u}$ in a continuum medium. 
    The square mesh indicates the atomic displacement described by the field ${\bm u}$, accompanied by a local rotation in the $x_1$-$x_2$ plane, whereas the ellipse denotes the rotation of the internal degree.
    Here, $a$ is a lattice constant and ${\bm e}_i$ ($i=1,2$) are unit vectors.}
    \label{fig:microrotation}
\end{figure}

The MPE originates in the Cosserat continuum~\cite{Cosserat1909} and was later systematized by Eringen as a microcontinuum field theory for media with internal rotations and microstructure~\cite{Eringen1966,Eringen1999}.
In a micropolar medium, the microrotation field ${\bm \phi}({\bm x})$ is introduced independently of the displacement field ${\bm u}({\bm x})$.
Figure~\ref{fig:microrotation} illustrates the physical meaning of these two fields.
The displacement field ${\bm u}$ represents the spatial displacement and associated rotation of the unit cell in solids.
In addition, MPE incorporates an internal rotational degree of freedom independently, representing a molecule or a rigid cluster with strong bonds inside a unit cell (denoted by the ellipse in Fig.~\ref{fig:microrotation}).
The rotation axis and angle are described by ${\bm \phi}$, which includes the rotation angle $\phi$ ($=\phi_3$) in Fig.~\ref{fig:microrotation}.
In general, the three-component microrotation field ${\bm \phi}$ represents the inertial rotational motion of the microstructure associated with optical phonon modes~\cite{Kishine2020}.

In the micropolar elasticity theory, the strain measures are described by two tensors, namely, the microdeformation tensor
\begin{align}
    \epsilon_{ij} = \partial_i u_j - \varepsilon_{ijk} \phi_k ,
\end{align}
and the wryness (microrotation) tensor
\begin{align}
    \gamma_{ij} = \partial_j \phi_i .
\end{align}
We note that if the microrotation field is fully locked to the lattice rotation as $\phi_i = \omega_i=\varepsilon_{ijk}\partial_j u_k/2$ (${\bm \phi}={\bm\omega}={\rm rot}\,{\bm u}/2$), the conventional elasticity theory is reproduced from Eq.~(\ref{epsilonijC}).
The strain energy density takes the quadratic form~\cite{Eringen1999,Nowacki1985}
\begin{align}
    U & = A_{ijkl} \epsilon_{ij} \epsilon_{kl}
    + B_{ijkl} \gamma_{ij} \gamma_{kl}
    + C_{ijkl} \epsilon_{ij} \gamma_{kl},
\end{align}
where $A_{ijkl}$, $B_{ijkl}$, and $C_{ijkl}$ are fourth-rank tensors.
The microdeformation tensor $\epsilon_{ij}$ is even under spatial inversion because $u_i$ and $\partial_j$ are polar vectors.
Therefore, $A_{ijkl}$ and $B_{ijkl}$ are both polar tensors and exist irrespective of inversion symmetry.
On the other hand, the wryness tensor $\gamma_{ij}$ is odd under inversion because ${\bm \phi}$ is an axial vector.
Thus, $C_{ijkl}$ is an axial tensor, which vanishes in systems with inversion symmetry.
In other words, the tensor $C_{ijkl}$ represents the effect of structural chirality in crystals.

\begin{figure}
    \centering    \includegraphics[width=0.8\columnwidth]{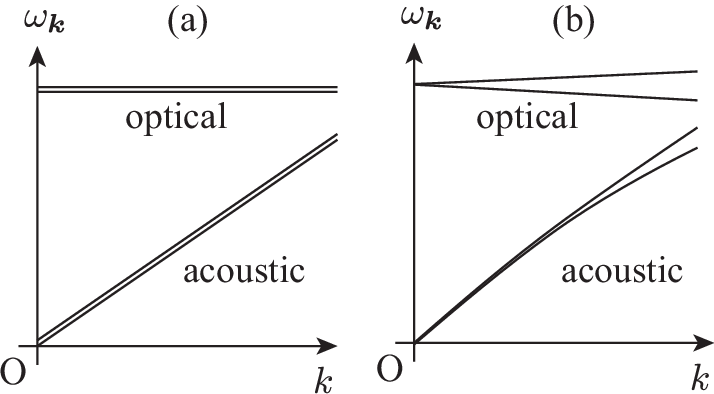}
    \caption{Schematic of the phonon dispersion relations in the long-wavelength limit for transverse phonons in (a) achiral materials and (b) chiral materials.
        A pair of adjacent lines in (a) indicates the two-fold degeneracy of phonon modes. The magnitude of the frequency splitting in (b) is proportional to $k$ for optical phonons and to $k^2$ for acoustic phonons.}
    \label{fig:dispersion_outline}
\end{figure}

We briefly sketch the transverse-phonon dispersions of isotropic micropolar elasticity, derived in Sec.~\ref{sec:local_frame_microcontinua}.
In achiral materials, the two circular polarizations remain degenerate, as shown in Fig.~\ref{fig:dispersion_outline}(a).
The acoustic and optical branches may still hybridize, while no left--right chiral splitting occurs.
On the other hand, in chiral materials ($C_{ijkl}\ne 0$), the chiral coupling lifts the left--right degeneracy of the two circular polarizations.
As a result, the two optical modes become nondegenerate at finite $k$ and exhibit dispersion relations of the form $\omega_{\bm k} = \omega_0 \pm c' k$, where $c'$ is a constant and the sign is determined by the phonon chirality.
The two acoustic modes become nondegenerate at order $k^2$, with the dispersion relations $\omega_{\bm k} = ck(1 \pm \lambda k)$, where $\lambda$ is a constant and the sign is determined by the phonon chirality.
These features, which are schematically shown in Fig.~\ref{fig:dispersion_outline}(b), are indeed observed in the phonon dispersion of real chiral materials~\cite{Hansen1980,Komiyama2022,Tsunetsugu2023,Kato2023,Tsunetsugu2026}.

Thus acoustic chiral phonons require coupling to the microrotation field, namely, the optical mode, through the axial tensor $C_{ijkl}$.
In a continuum model retaining only acoustic phonons, structural chirality is effectively erased and must be restored through coupling to microrotation.

\subsection{Local \texorpdfstring{$\mathrm{SO}(3)$}{SO(3)} frame formulation}

The local rotational degrees of freedom needed for optical phonons are represented by a material frame attached to the microstructure. The local-frame formulation is used here to make the angular momentum carried by acoustic and optical phonons explicit, rather than to present micropolar elasticity as a new theory.

In the local frame formulation, we introduce a local material frame $Q^a_{\ i}(\bm{x})\in \mathrm{SO}(3)$. 
This frame describes the rotation of the internal degree of freedom, which can be intuitively regarded as a rigid ellipsoid in a unit cell (see Fig.~\ref{fig:microrotation}).
More precisely, $Q^a_{\ i}(\bm{x})$ represents the transformation from the Euclidean frame to the co-rotating frame associated with the ellipsoid.
Here, the Latin indices ($a,b,\cdots =1,2,3$) denote spatial components in the three-dimensional co-rotating frame.
The position of the unit cell under lattice displacement is $y^i(t,{\bm x}) = x^i + u^i(t,{\bm x})$, and the deformation gradient is $F^i_{\ j}=\partial_j y^i = \delta^i_{\ j} + \partial_j u^i$.
We define the co-rotated deformation gradient in the co-rotating frame as
\begin{align}
    E^a_{\ j}=(Q^{-1})^a_{\ i}F^i_{\ j} .
\end{align}
This quantity contains information about both the coordinate rotation (i.e., the rotation of the square mesh in Fig.~\ref{fig:microrotation}) and the rotation of the ellipsoid.
On the other hand, the rotational connection associated with the spatial variation of the local frame,
\begin{align}
    \Omega_i^{ab}=(Q^{-1}\partial_i Q)^{ab}
\end{align}
measures how the orientation of the ellipsoid changes from point to point and is related to the rotation vector ${\bm \phi}$ (for details, see Sec.~\ref{sec:eringen_reduction}).

Here the material frame is fixed by the microstructure. It is therefore a physical local frame used to describe rotational optical degrees of freedom, not an additional redundancy of the elastic variables.

Within the local $\mathrm{SO}(3)$ frame formulation, the following points become clear:
\begin{enumerate}
    \item \textbf{Unified phonon angular momentum:} The total angular momentum follows from Noether's theorem as the sum of displacement polarization and intrinsic microrotation. This establishes a continuum definition of phonon angular momentum, rather than only a change of variables in micropolar elasticity.
    \item \textbf{Vorticity in conservation laws:} Lattice vorticity enters as a locking limit, $\boldsymbol{\phi} \approx \nabla \times \mathbf{u}/2$, rather than as a fundamental field. The resulting conservation law reduces to the vorticity-driven form used for surface acoustic waves.
    \item \textbf{Generalized continua and chirality:} The same variables extend to micromorphic, microstretch, and micropolar continua, and they connect parity-breaking terms  to the local frame.
    \item \textbf{Phonon dispersion:} The dispersion relations of acoustic and optical phonons can be obtained analytically. In the presence of a symmetry-allowed inversion- and mirror-symmetry-breaking term, the formulation captures chiral splitting of the phonon dispersion.
\end{enumerate}
\subsection{Relation to earlier local-frame formulations}

We briefly recall earlier local-frame formulations of micropolar and micromorphic continua.
Riemann--Cartan treatments of gravity identified torsion and curvature as independent geometric fields associated with spin and moment stresses~\cite{Hehl1976}, and they motivated continuum descriptions of solids with defects and microstructure, where dislocations and disclinations appear as field strengths, and nonsymmetric stresses and defect densities are treated on the same footing~\cite{Kleinert1989,Lagoudas1989,Lazar2009,lazar2009PhysicsLettersA}.
Local-frame formulations closer to Eringen interpret the wryness tensor as a connection and relate generalized continua to defect variables and microdistortions~\cite{GayBalmazRatiu2010,neff2015relaxed,BoehmerObukhov2012}.
The present work uses the local material frame for a different purpose: to formulate the long-wavelength phonon sector in which acoustic displacements and optical rotations carry angular momentum within the same continuum theory.
The formulation identifies the rotational variables relevant to electron-spin coupling.

\section{Local \texorpdfstring{$\mathrm{SO}(3)$}{SO(3)} material-frame formulation}
\label{sec:local_frame_microcontinua}

We now formulate the local rotational sector of Eringen microcontinua directly in terms of macroscopic deformation and a material-frame microrotation.
The key step is to introduce co-rotated variables built from the local material frame and then to show that, in the linear long-wavelength rotational sector, the standard micropolar variables emerge.
Readers primarily interested in the reduction of the present formulation to linear MPE may start with Sec.~\ref{sec:mpe_linear} and refer back to Sec.~\ref{sec:eringen_reduction} as needed.

\subsection{Matter-fixed parametrization by macroscopic deformation and microrotation}
\label{sec:eringen_reduction}

We now connect the local-frame variables to standard elastic degrees of freedom.
The macroscopic configuration is described by an embedding $y^i(t,\bm{x})$ of the material point $\bm{x}$ into physical space.
The deformation gradient is
\begin{align}
    F^i_{\ j}(t,\bm{x})=\partial_j y^i(t,\bm{x}).
\end{align}
The independent microstructural rotation is represented by a microrotation matrix
\begin{align}
    Q^a_{\ i}(t,\bm{x})\in \mathrm{SO}(3).
\end{align}
The index $a$ labels components in the material frame attached to the microstructure, whereas $i,j$ label spatial components.
The orientation $Q$ is fixed by the microstructure and is treated as a physical microrotation field, not as a gauge redundancy.

To separate macroscopic deformation from internal rotation, we introduce a relative deformation measured in the local material frame, defined by
\begin{align}
    E^a_{\ j}\equiv (Q^{-1})^a_{\ i}\,F^i_{\ j}.
    \label{eq:E_def}
\end{align}
$E^a_{\ j}$ measures the deformation gradient expressed in the local microstructural frame.
The local-frame covariant measure of rotational gradients is the rotational connection associated with $Q$,
\begin{align}
    \Omega_{\mu}^{ab}\equiv (Q^{-1}\partial_{\mu}Q)^{ab}.
    \label{eq:Omega_def}
\end{align}
Its axial-vector representations are
\begin{align}
    \Omega_k & \equiv -\frac{1}{2}\varepsilon_{abk}\,\Omega_{0}^{ab},
    \label{eq:axial_defs}
    \\
    W_{ik}  & \equiv -\frac{1}{2}\varepsilon_{abk}\,\Omega_{i}^{ab}.
    \label{eq:Wik}
\end{align}
The minus sign fixes the convention in which the axial vector $\bm{\phi}$ used below coincides with the standard micropolar microrotation.
$\bm{\Omega}$ represents the angular velocity of the microrotation, while $W_{ik}$ describes its spatial microrotation gradient.

\subsection{Rotational sector within the microdistortion hierarchy}

We summarize how the present rotational sector fits into Eringen's broader microcontinuum hierarchy. Replacing the microrotation $Q^a_{\ i}(t,\bm{x})\in \mathrm{SO}(3)$ by a general microdistortion $P^a_{\ i}(t,\bm{x})\in \mathrm{GL}(3)$ yields
\begin{align}
    E^a_{\ j} & \equiv (P^{-1})^a_{\ i}\,F^i_{\ j},       \\
    \Omega_{\mu}{}^{a}{}_{b} & \equiv (P^{-1}\partial_{\mu}P)^{a}{}_{b}.
\end{align}
The $\mathrm{GL}(3)$ connection $\Omega_{\mu}$ can be decomposed into an antisymmetric part (microrotation), a trace part (microstretch), and a symmetric traceless part (microshear).
Micropolar elasticity is recovered by the restriction $P=Q\in \mathrm{SO}(3)$.
Microstretch models can be obtained by $P=\zeta Q$ with a scalar stretch field $\zeta>0$ and $Q\in \mathrm{SO}(3)$.
Intuitively, microstretch models describe a change in the size of the microstructural object (the ellipsoid).
Micromorphic models correspond to retaining all nine components of $P$, which include microshear fields for describing quadrupolar-type deformations of the ellipsoid without a change in size.
Thus, the correspondence is direct: the three antisymmetric components describe micropolar microrotation; adding the scalar trace yields the $1\oplus 3$ sector of microstretch theory; and including the five symmetric traceless components yields the full $1\oplus 3\oplus 5$ micromorphic microdistortion.

Since the optical phonon modes considered here are associated with rotational degrees of freedom, the micropolar restriction $P\in \mathrm{SO}(3)$ selects the rotational low-energy sector needed for the angular-momentum problem.
Below we restrict attention to the micropolar specialization $P=Q\in \mathrm{SO}(3)$, for which $\Omega_{\mu}^{ab}$ is $\mathfrak{so}(3)$-valued and the axial representations Eqs.~\eqref{eq:axial_defs} and \eqref{eq:Wik} apply.

\subsection{Linear reduction to microdeformation and wryness}
\label{sec:linearization_epsilon_gamma}

We linearize around a reference configuration $y^i=x^i$ and $Q=\mathbf{1}$ by writing
\begin{align}
    y^i(t,\bm{x})     & =x^i+u^i(t,\bm{x}),                                  \\
    Q(t,\bm{x}) & =\exp\!\big[-(\bm{\phi}(t,\bm{x})\times)\big],
\end{align}
where $(\bm{\phi}\times)^a_{\ i}=\varepsilon_{aik} \phi_k$ is the antisymmetric matrix generated by the axial vector $\bm{\phi}$, with the sign chosen to make the linearized strain consistent with the Eringen--Cosserat convention.
To first order,
\begin{align}
    F^i_{\ j}              & =\delta^i_{\ j}+\partial_j u^i,                \\
    Q^a_{\ i}        & =\delta^a_{\ i}-\varepsilon_{aik}\phi_k, \\
    (Q^{-1})^a_{\ i} & =\delta^a_{\ i}+\varepsilon_{aik}\phi_k.
\end{align}
Substitution into Eq.~\eqref{eq:E_def} yields
\begin{align}
    E^a_{\ j}
     & =(Q^{-1})^a_{\ i}F^i_{\ j}\notag                                                        \\
     & =\delta^a_{\ j}+\partial_j u^a+\varepsilon_{ajk}\phi_k+\mathcal{O}(u\phi,\phi^2).
\end{align}
We define the microdeformation tensor from the co-rotated deformation gradient as
\begin{align}
\epsilon_{ij} & \equiv \delta_{aj}\left(E^a_{\ i}-\delta^a_{\ i}\right).
\end{align}
Using Eq.~\eqref{eq:E_def} to linear order, this becomes
\begin{align}
\epsilon_{ij}=\partial_i u_j-\varepsilon_{ijk}\phi_k.
    \label{eq:epsilon_Eringen}
\end{align}
This expression corresponds to the standard Eringen micropolar microdeformation tensor.
The microrotation enters at linear order because $E=Q^{-1}F$ compares the macroscopic deformation gradient with the local microstructural orientation.

Next, linearizing Eq.~\eqref{eq:Omega_def} yields
\begin{align}
    \Omega_i^{ab}
     & =(Q^{-1}\partial_i Q)^{ab}\notag \\
     & =-\varepsilon_{abk}\,\partial_i\phi_k+\mathcal{O}(\phi\,\partial\phi).
\end{align}
Therefore,
\begin{align}
W_{ik}=\partial_i\phi_k+\mathcal{O}(\phi\,\partial\phi).
\end{align}
In the linear micropolar model, we identify the wryness tensor as
\begin{align}
    \gamma_{ij}\equiv \partial_j\phi_i.
    \label{eq:gamma_Eringen}
\end{align}
$\gamma_{ij}$ is associated directly with the rotational connection itself, i.e., with the first spatial gradient of the local material frame, rather than with a higher-derivative field strength.
Such higher-derivative terms appear only beyond leading order in the long-wavelength expansion when $\Omega_i^{ab}$ is induced from $\phi$ (see Sec.~\ref{sec:eringen_reduction}).

\subsection{Quadratic energy in the linear micropolar reduction}
\label{sec:mpe_linear}

The linear isotropic MPE Lagrangian is constructed from $\epsilon_{ij}$ and $\gamma_{ij}$.
We decompose a generic second-rank tensor $X_{ij}$ as
\begin{align}
    X_{ij}   & =X_{(ij)}+X_{[ij]},          \\
    X_{(ij)} & =\frac{1}{2}(X_{ij}+X_{ji}), \\
    X_{[ij]} & =\frac{1}{2}(X_{ij}-X_{ji}),
\end{align}
The trace is denoted by $X_{kk}$.
For an isotropic nonchiral medium, the parity-even quadratic energy in the strain and wryness sectors can be parametrized by three moduli in each sector:
\begin{align}
    U_{\text{strain}}(\epsilon)
     & =\frac{\lambda}{2}(\epsilon_{kk})^2
    +\mu\,\epsilon_{(ij)}\epsilon_{(ij)}
    +\kappa\,\epsilon_{[ij]}\epsilon_{[ij]},
    \label{eq:U_strain}                    \\
    U_{\text{wryness}}(\gamma)
     & =\frac{\alpha}{2}(\gamma_{kk})^2
    +\frac{\beta}{2}\,\gamma_{(ij)}\gamma_{(ij)}
    +\frac{\gamma}{2}\,\gamma_{[ij]}\gamma_{[ij]}.
    \label{eq:U_wryness}
\end{align}
This form is equivalent to assigning independent coefficients to the $\mathrm{SO}(3)$ irreducible components $1\oplus 3\oplus 5$, after decomposing $\epsilon_{(ij)}\epsilon_{(ij)}$ and $\gamma_{(ij)}\gamma_{(ij)}$ into their trace and symmetric-traceless parts.
More explicitly, at the level of linear second-rank tensor variables, the nine components decompose under spatial $\mathrm{SO}(3)$ rotations into a scalar trace $1$, an antisymmetric rotational part $3$, and a symmetric-traceless part $5$.
This decomposition matches the usual hierarchy of generalized continua: the rotational micropolar sector keeps the $3$, microstretch theory keeps $1\oplus3$, and the full micromorphic theory keeps $1\oplus3\oplus5$.

The kinetic energy density is
\begin{align}
    \mathcal{K}
    =\frac{\rho}{2}\dot{u}_i\dot{u}_i+\frac{I}{2}\dot{\phi}_i\dot{\phi}_i,
\end{align}
which is the small-angle reduction of $(I/2)\Omega_i\Omega_i$ with $\bm{\Omega}\simeq\dot{\bm{\phi}}$.
The linear MPE Lagrangian density is then
\begin{align}
    \mathcal{L}_{\text{MPE}}
    =\mathcal{K}-U_{\text{strain}}(\epsilon)-U_{\text{wryness}}(\gamma),
    \label{eq:L_MPE}
\end{align}
with $\epsilon_{ij}$ and $\gamma_{ij}$ given by Eqs.~\eqref{eq:epsilon_Eringen} and \eqref{eq:gamma_Eringen}.

It is useful to make explicit how the antisymmetric strain sector reduces to the standard Cosserat relative-rotation term.
From Eq.~\eqref{eq:epsilon_Eringen},
\begin{align}
    \epsilon_{[ij]}
     & =\frac{1}{2}(\partial_i u_j-\partial_j u_i)-\varepsilon_{ijk}\phi_k.
    \label{eq:eps_asym_def}
\end{align}
Introducing the axial vector $A_k\equiv \frac{1}{2}\varepsilon_{kij}\epsilon_{[ij]}$ yields
\begin{align}
    A_k=\frac{1}{2}(\nabla\times\bm{u})_k-\phi_k,
\end{align}
and therefore
\begin{align}
    \epsilon_{[ij]}\epsilon_{[ij]}
    =2A_kA_k
    =\frac{1}{2}\left|\nabla\times\bm{u}-2\bm{\phi}\right|^2.
    \label{eq:cosserat_identity}
\end{align}
The Cosserat modulus $\kappa$ penalizes the mismatch between the macroscopic vorticity and the microrotation.

\subsection{Euler--Lagrange equations, stress, and couple stress}

We define the force-stress tensor and couple-stress tensor by
\begin{align}
    \sigma_{ij} & \equiv \frac{\partial U}{\partial \epsilon_{ij}}, \\
    m_{ij}      & \equiv \frac{\partial U}{\partial \gamma_{ij}},
\end{align}
with $U=U_{\text{strain}}+U_{\text{wryness}}$.
For the isotropic quadratic form given in Eqs.~\eqref{eq:U_strain} and \eqref{eq:U_wryness},
\begin{align}
    \sigma_{ij}
     & =\lambda\,\epsilon_{kk}\delta_{ij}
    +2\mu\,\epsilon_{(ij)}
    +2\kappa\,\epsilon_{[ij]},
    \label{eq:sigma_constitutive}         \\
    m_{ij}
     & =\alpha\,\gamma_{kk}\delta_{ij}
    +\beta\,\gamma_{(ij)}
    +\gamma\,\gamma_{[ij]}.
    \label{eq:m_constitutive}
\end{align}
The Euler--Lagrange equations derived from Eq.~\eqref{eq:L_MPE} are
\begin{align}
    \rho\,\ddot{u}_i & = \partial_j\sigma_{ji},
    \label{eq:EOM_u}                                                    \\
    I\,\ddot{\phi}_i & =\varepsilon_{ijk}\sigma_{jk}+\partial_j m_{ij}.
    \label{eq:EOM_phi}
\end{align}
Equation~\eqref{eq:EOM_phi} is the linear balance equation for the intrinsic microrotation angular momentum: the antisymmetric part of the force stress supplies a local torque, while $\partial_j m_{ij}$ is the divergence of the couple stress.

\subsection{Global rotations and angular momentum in the linear micropolar reduction}
\label{sec:noether_global_rotation_mpe}

Global rotational invariance leads to the angular momentum conservation law stated below, together with the explicit decomposition of the angular momentum density used later.

Consider a rigid rotation with constant $R_0=\exp[(\bm{\Theta}\times)]\in \mathrm{SO}(3)$.
At fixed material coordinate $\bm{x}$, the embedding field and microrotation transform as
\begin{align}
    \bm{y}(t,\bm{x}) & \to \bm{y}'(t,\bm{x})=R_0\,\bm{y}(t,\bm{x}), \\
    Q(t,\bm{x})      & \to Q'(t,\bm{x})=R_0\,Q(t,\bm{x}).
\end{align}
This transformation is a physical global rotation of the embedded body and its microstructure.\footnote{Here the rotation acts on the physical spatial orientation of both the macroscopic body and the local material frame attached to the microstructure. This physical rotation differs from a mere relabeling of material-frame components, which would change the local basis used to describe the same physical configuration. The present formulation uses $Q$ as a physical microrotation field rather than as a gauge redundancy.}
A Lagrangian constructed from $E=Q^{-1}F$ and $\Omega_{\mu}=Q^{-1}\partial_{\mu}Q$ is invariant under such a rigid rotation.
Noether's first theorem then implies the local conservation law
\begin{align}
    \partial_t J_k^{0}+\partial_i J_k^{i}=0.
    \label{eq:Noether_AM}
\end{align}

The corresponding angular momentum density can be written in terms of the momentum density $\bm{p}$ and the microrotation angular velocity $\bm{\Omega}$ as
\begin{align}
    J_k^{0}
    =
    \varepsilon_{kij}y_i p_j
    +
    \big(I\bm{\Omega}\big)_k,
    \label{eq:J0_general}
\end{align}
with $\bm{p}=\rho\,\dot{\bm{y}}$ and $\bm{\Omega}$ defined in Eq.~\eqref{eq:axial_defs}.
In linear MPE, $\bm{y}=\bm{x}+\bm{u}$ and $\dot{\bm{y}}=\dot{\bm{u}}$; hence Eq.~\eqref{eq:J0_general} decomposes as
\begin{align}
    J_k^{0}
    &=
    \big(\bm{x}\times\rho\dot{\bm{u}}\big)_k
    +
    \big(\bm{u}\times\rho\dot{\bm{u}}\big)_k
    +
    \big(I\bm{\Omega}\big)_k \notag \\
    &\simeq
    \big(\bm{x}\times\rho\dot{\bm{u}}\big)_k
    +
    \big(\bm{u}\times\rho\dot{\bm{u}}\big)_k
    +
    \big(I\dot{\bm{\phi}}\big)_k,
    \label{eq:J0_decomposition_linearphi}
\end{align}
where we have used the small-angle approximation $\bm{\Omega}\simeq\dot{\bm{\phi}}$.
Equation~\eqref{eq:J0_decomposition_linearphi} separates the total Noether angular momentum density into the macroscopic orbital part, the displacement-polarization part conventionally used for phonons, and the intrinsic contribution carried by the local material rotation.~\footnote{The second term in Eq.~\eqref{eq:J0_decomposition_linearphi} arises from the decomposition $\bm{y}=\bm{x}+\bm{u}$ in the linearized regime, and it is useful to keep it explicit when comparing different conventions in the phonon angular momentum literature. The microrotation contribution is expressed as $I\bm{\Omega}$, with components $\Omega_k=-\varepsilon_{abk}(Q^{-1}\partial_t Q)^{ab}/2$ in the convention used here. The replacement $\bm{\Omega}\simeq\dot{\bm{\phi}}$ corresponds to the small-angle approximation. Treating $\bm{\phi}$ as an ordinary vector with a homogeneous variation under rotations leads to a different canonical expression that does not reproduce the physical rotor angular momentum.}

\subsection{Vorticity form of internal angular momentum in the strong-coupling limit}
\label{sec:locking_vorticity_AM}

Consider next the locking limit in which the microrotation follows the local lattice rotation.
Taking the time derivative of this locking relation then connects the microrotation angular velocity to the vorticity of the lattice velocity field.
This limit connects MPE to the vorticity-based descriptions used for surface acoustic waves.

The Cosserat contribution to the potential energy is
\begin{align}
    U_{\mathrm{C}}
    =
    \kappa\,\epsilon_{[ij]}\epsilon_{[ij]}
    =
    \frac{\kappa}{2}\left|\nabla\times\bm{u}-2\bm{\phi}\right|^2,
    \label{eq:U_Cosserat}
\end{align}
using Eq.~\eqref{eq:cosserat_identity}.
In the limit $\kappa\to\infty$, configurations with finite energy satisfy the locking constraint
\begin{align}
    \epsilon_{[ij]} & =0,     \quad \Rightarrow \quad \phi_k=\frac{1}{2}\varepsilon_{kij}\partial_i u_j.
    \label{eq:locking_constraint_phi_curlu}
\end{align}
Taking the time derivative yields
\begin{align}
    \dot{\phi}_k
    =
    \frac{1}{2}\varepsilon_{kij}\partial_i \dot{u}_j
    \equiv
    \varpi_k,
    \label{eq:locking_constraint_phidot_vorticity}
\end{align}
where $\bm{\varpi} =\nabla\times \dot{\bm u}/2$ is the local vorticity of the lattice velocity field $\dot{\bm{u}}$.

In linear MPE, the internal angular momentum density carried by microrotation is $I\dot{\bm{\phi}}$ in Eq.~\eqref{eq:J0_decomposition_linearphi}.
Using Eq.~\eqref{eq:locking_constraint_phidot_vorticity}, it reduces to
\begin{align}
    \big(I\dot{\bm{\phi}}\big)_k
    \to
    I\varpi_k
    =
    \frac{I}{2}\left(\nabla\times\dot{\bm{u}}\right)_k.
    \label{eq:microspin_to_vorticity}
\end{align}
Substitution into Eq.~\eqref{eq:J0_decomposition_linearphi} yields, in the locking limit,
\begin{align}
    J_k^{0}
    \to
    \varepsilon_{kij}x_i\big(\rho\dot{u}_j\big)
    +
    \varepsilon_{kij}u_i\big(\rho\dot{u}_j\big)
    +
    \frac{I}{2}\varepsilon_{kij}\partial_i\dot{u}_j.
    \label{eq:J0_locking_limit}
\end{align}
If $I$ is spatially uniform, the vorticity term can be written as a total derivative,
\begin{align}
    \frac{I}{2}\varepsilon_{kij}\partial_i\dot{u}_j
    =
    \partial_i\left(\frac{I}{2}\varepsilon_{kij}\dot{u}_j\right).
    \label{eq:vorticity_total_derivative}
\end{align}
Therefore, for spatially uniform $I$, the volume integral of the intrinsic vorticity contribution is determined by boundary terms and vanishes under appropriate boundary conditions.
For large, finite $\kappa$, $\bm{\phi}$ remains dynamical and is not strictly equal to $\nabla\times\bm{u}/2$, and the relative rotation $\nabla\times\bm{u}/2-\bm{\phi}$ constitutes a genuine dynamical variable.

\subsection{Phonon dispersions: acoustic and optical branches}
\label{sec:dispersion}

\begin{figure}
    \centering
    \includegraphics[width=\columnwidth]{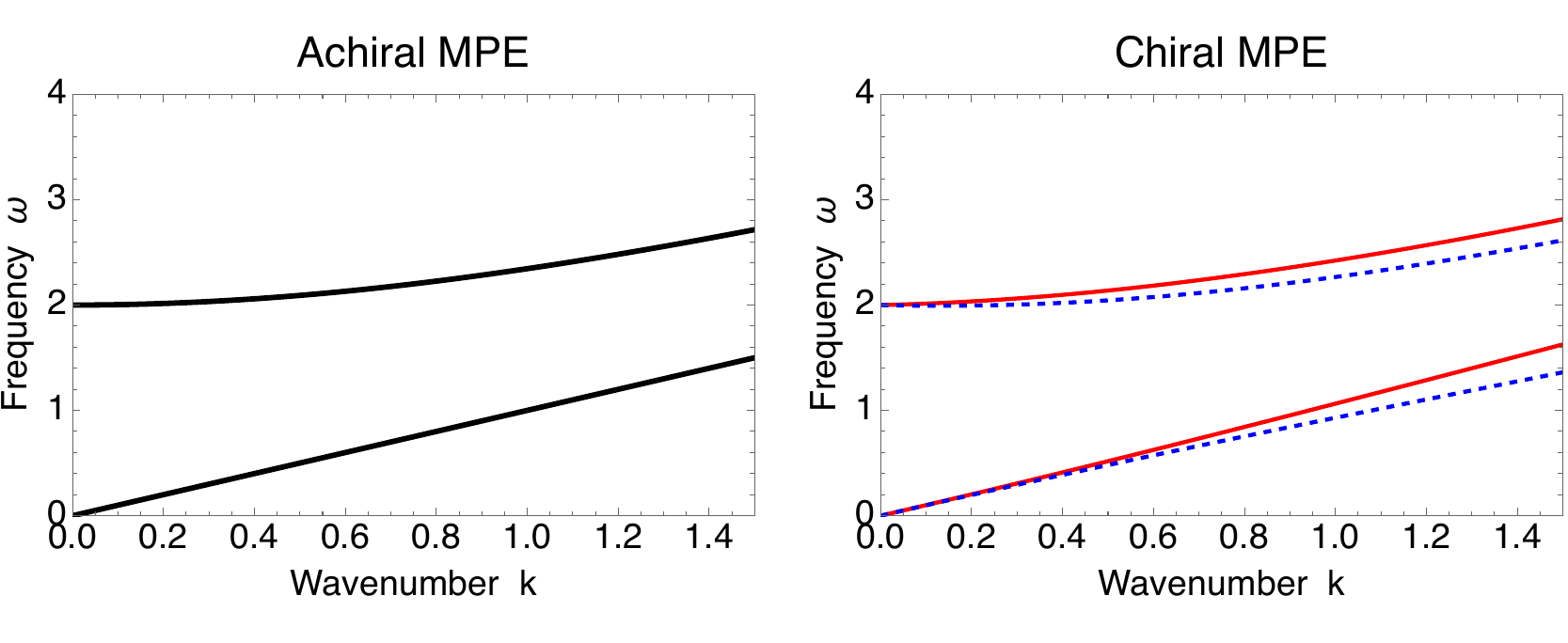}
    \caption{Transverse phonon dispersions in micropolar elasticity. In the achiral case, the acoustic and optical branches remain degenerate, whereas in the chiral case, the optical branch is split and the acoustic branch exhibits a weaker splitting. The curves are plotted for representative dimensionless parameters $\rho=1$, $I=0.5$, $\mu=1$, $\kappa=0.5$, and $\beta+\gamma=1$. The achiral panel corresponds to $\chi=0$, whereas the chiral panel uses $\chi=0.2$ in the convention of Eq.~\eqref{eq:chiral_term}. No material-specific fit is implied.}
    \label{fig:dispersion_mpe_section}
\end{figure}

Within the linear micropolar reduction, the phonon dispersion relations can be obtained analytically.
Plane-wave solutions of Eqs.~\eqref{eq:sigma_constitutive}--\eqref{eq:EOM_phi} take the form $u_i,\phi_i\propto \exp[i(\bm{k}\cdot\bm{x}-\omega t)]$.
For the longitudinal sector, the displacement component parallel to $\bm{k}$ decouples from the microrotation, and one obtains the longitudinal acoustic branch
\begin{align}
    \omega_L^2 & =c_L^2 k^2,                 \\
    c_L^2      & =\frac{\lambda+2\mu}{\rho}.
    \label{eq:disp_longitudinal}
\end{align}
In the transverse sector, the displacement and microrotation are coupled through the Cosserat term Eq.~\eqref{eq:cosserat_identity}.
For $\bm{k}=k\hat{\bm{z}}$, each of the two degenerate transverse polarizations reduces to an equivalent $2\times 2$ eigenvalue problem and yields two branches
\begin{align}
    \omega_{\pm}^2(k)
     & =\frac{1}{2}\left(\frac{\mu+\kappa}{\rho}k^2+\frac{4\kappa}{I}+\frac{\beta+\gamma}{2I}k^2\right)\notag \\
     & \quad \pm
    \frac{1}{2}\sqrt{\left(\frac{\mu+\kappa}{\rho}k^2-\frac{4\kappa}{I}-\frac{\beta+\gamma}{2I}k^2\right)^2+\frac{16\kappa^2}{\rho I}k^2}.
    \label{eq:disp_transverse_mixing}
\end{align}
At small $k$, $\omega_{-}(k)\simeq c_T k$ corresponds to the transverse acoustic branch with $c_T^2=\mu/\rho$, while $\omega_{+}(k)$ is an optical branch with a gap
\begin{align}
    \omega_0^2=\frac{4\kappa}{I},
    \label{eq:optical_gap}
\end{align}
which arises from mode mixing between transverse vorticity and microrotation encoded in the Cosserat sector.
The remaining microrotation component parallel to $\bm{k}$ is likewise gapped; in the present isotropic parametrization its dispersion is $\omega_\parallel^2=4\kappa/I+(\alpha+\beta)k^2/I$.
Figure~\ref{fig:dispersion_mpe_section} summarizes the corresponding transverse dispersions in achiral and chiral MPE.

\subsection{Chiral splitting of circularly polarized modes}

The achiral isotropic quadratic energy in Eqs.~\eqref{eq:U_strain} and \eqref{eq:U_wryness} is invariant under improper spatial operations such as spatial inversion and mirror reflection.
In chiral media, where such improper symmetries are absent, pseudoscalar terms that remain invariant under proper rotations are allowed.
A minimal and symmetry-allowed example in the microrotation sector is~\footnote{The term in Eq.~\eqref{eq:chiral_term} is the bulk representative of a chiral micropolar coupling. Indeed,
$\bm{\phi} \cdot (\nabla\times\bm{\phi}) = \epsilon_{ij}\gamma_{ij}-\epsilon_{ii}\gamma_{kk}-\partial_i(u_j\partial_j\phi_i-u_i\partial_j\phi_j)$,
which is equivalent, modulo a surface term, to $C_{ijkl}\epsilon_{ij}\gamma_{kl}$ with $C_{ijkl}=(\chi/2)(\delta_{ik}\delta_{jl}-\delta_{ij}\delta_{kl})$.}
\begin{align}
    U_{\chi}
    =
    \frac{1}{2}
    \chi\,\varepsilon_{ijk}\,\phi_i\,\partial_j\phi_k,
    \label{eq:chiral_term}
\end{align}
with a material-dependent coefficient $\chi$.
For $\bm{k}=k\hat{\bm{z}}$, this term lifts the degeneracy between left and right circularly polarized transverse microrotation waves.
In the regime dominated by the optical scale, one finds the characteristic linear-in-$k$ splitting
\begin{align}
    \omega_{\mathrm R/\mathrm L}^2(k)=\omega_0^2+c_{\phi}^2 k^2\pm \frac{\chi}{I}k+\cdots,
    \label{eq:chiral_splitting}
\end{align}
where $c_{\phi}$ denotes the achiral $k^2$ stiffness of the corresponding predominantly microrotational optical branch; its detailed value depends on the wryness moduli and on displacement--microrotation mixing.
This linear-in-$k$ splitting is a direct consequence of inversion- and mirror-symmetry-breaking microrotation dynamics.
These features are illustrated in Fig.~\ref{fig:dispersion_mpe_section} using representative parameters, for which no material-specific fitting is implied (for qualitative discussion, see Sec.~\ref{MPEtheory}).
Thus the same displacement--microrotation coupling that defines the Noether angular momentum transfers chiral microrotation splitting to the acoustic branch through mode hybridization.

\section{Finite microrotations and angular momentum beyond the linear approximation}
\label{sec:nonlinear_Q_model}

The continuum angular-momentum density obtained above can be extended beyond the linear micropolar approximation by keeping the microrotation as a finite $\mathrm{SO}(3)$ field $Q(t,\bm{x})$, rather than expanding $Q$ to first order in the rotation coordinate~\footnote{No torsion or curvature terms are required for this discussion.}.

Retaining the co-rotated deformation gradient $E=Q^{-1}F$ and the rotational connection, the rotational kinetic energy is written as
\begin{align}
\mathcal{K}_{\text{rot}} & = \frac{I}{2}\Omega_i\Omega_i, \\
\Omega_i &= -\frac{1}{2}\varepsilon_{abi}\,(Q^{-1}\partial_t Q)^{ab}.
    \label{eq:Krot_nonlinear}
\end{align}
The internal angular momentum density carried by the microrotation is
\begin{align}
    S_i^{\text{micro}}=I\,\Omega_i.
    \label{eq:micro_spin_exact}
\end{align}
If $Q=\exp[-(\bm{\phi}\times)]$ is parameterized by a rotation vector $\bm{\phi}$, the relation between $\bm{\Omega}$ and $\dot{\bm{\phi}}$ is nonlinear:
\begin{align}
    \bm{\Omega}=J(\bm{\phi})\,\dot{\bm{\phi}},
\end{align}
where $J(\bm{\phi})$ is the $\mathrm{SO}(3)$ Jacobian written with the matrix convention $(\bm{\phi}\times)^a{}_{i}=\varepsilon_{aik}\phi_k$ used above, with a closed form
\begin{align}
    J(\bm{\phi})
           & =
    \mathbf{1}
    +\frac{1-\cos\theta}{\theta^2}(\bm{\phi}\times)
    +\frac{\theta-\sin\theta}{\theta^3}(\bm{\phi}\times)^2, \\
    \theta & =\lvert\bm{\phi}\rvert.
\end{align}
For small $\bm{\phi}$, this implies
\begin{align}
    \bm{\Omega}
    \simeq
    \dot{\bm{\phi}}
    -\frac{1}{2}\,\bm{\phi}\times\dot{\bm{\phi}}
    +\mathcal{O}(\phi^2\dot{\phi}).
\end{align}
Thus the commonly used linear relation $S_i^{\text{micro}}\simeq I\dot{\phi}_i$ is the small-angle limit of a more general angular-momentum density defined by $I\bm{\Omega}$.
Therefore, the exact microrotation contribution in Eq.~\eqref{eq:micro_spin_exact} contains nonlinear corrections beyond the linear approximation $S_i^{\text{micro}}\simeq I\dot{\phi}_i$.

For global rigid rotations acting as $\bm{y}\to R_0\bm{y}$ and $Q\to R_0 Q$, Noether's first theorem yields a conserved angular momentum current satisfying Eq.~\eqref{eq:Noether_AM}.
The angular momentum density includes the exact microrotation contribution in Eq.~\eqref{eq:micro_spin_exact}.
Thus finite local rotations modify the constitutive expression for the angular-momentum density through $\bm{\Omega}(Q,\dot Q)$, while the Noether conservation law itself retains the same first-order continuity form.

This nonlinear extension is not used in the linear dispersion analysis above. It is relevant to soft optical modes involving finite local rotations, such as rigid octahedral rotations in perovskites considered in Ref.~\cite{Kishine2020}. Applications to realistic nonlinear lattice dynamics are left for future work.

\section{Conclusion}

We have formulated a continuum definition of phonon angular momentum in crystals by introducing a local $\mathrm{SO}(3)$ material frame that describes rotational optical degrees of freedom together with the macroscopic displacement field.
In the linear long-wavelength rotational sector, the construction reproduces the microdeformation and wryness tensors and contains isotropic micropolar elasticity as a special case.
Rotational symmetry and Noether's theorem then determine a definite angular-momentum density consisting of the displacement-polarization contribution and the intrinsic contribution from the local material rotation.
The same formulation shows how the vorticity-based description emerges in the strong-coupling limit and how improper-symmetry-breaking terms split chiral phonons.
In this sense, the present theory establishes a consistent continuum definition of phonon angular momentum applicable to both acoustic and optical modes.

This work has focused on local material rotations relevant to phonon angular momentum. The same local-frame construction can be extended to microdistortion fields that include internal stretch and shear degrees of freedom.
In the linearized limit, these sectors reproduce the broader Eringen hierarchy, including microstretch and micromorphic continua, and extend the continuum description to optical modes beyond purely rotational ones.
A more material-specific theory can further incorporate the crystallographic point group and the corresponding irreducible representations of optical modes.

The local material frame identifies the rotational variables that couple to electron spin in a low-energy theory.
This identification connects the present continuum variables to long-wavelength angular-momentum transfer between lattice degrees of freedom and electron spins~\cite{Matsuo2011,Matsuo2013,Wallis-APL-2006-09,Zolfagharkhani2008,ono2015Barnett,ogata2017Gyroscopic,hirohata2018magneto,imai2018observation,imai2019angular,harii2015Line,chudo2021Barnett,Kobayashi2017,kurimune2020Highly,kurimune2020Observation,tateno2020Electrical,tateno2021Einstein,Harii-2019-NatCommun,horaguchi2025nanometer,takahashi2016Spin,takahashi2020Giant,tabaeikazerooni2020Electron,tabaeikazerooni2021Electrical,tokoro2022spin}.
A detailed treatment of the electron-spin sector, including spin-rotation coupling and gyromagnetic effects in the presence of optical phonons, will be discussed in a separate paper.

\begin{acknowledgments}
    We especially thank J. Kishine for lectures and discussions on Eringen's micropolar theory and phonon angular momentum.
    We thank T. Funato, H. Funaki, R. Sano, Y. Sekino, and H. Tajima for useful discussions.
    This work was supported by the National Natural Science Foundation of China (NSFC) under Grant No. 12374126,
    by the Priority Program of Chinese Academy of Sciences under Grant No. XDB28000000,
    and by JSPS KAKENHI for Grants (Nos. 23H01839, 24H00322, 24K06951) from MEXT, Japan.
\end{acknowledgments}

\bibliography{ref}

\end{document}